\documentclass[titlepage,12pt]{article}
\usepackage{amssymb,amsmath,color,graphics,amscd,epsf,indentfirst,amsfonts}
\usepackage{enumerate}
\usepackage{epsfig}
\usepackage{hyperref}
\usepackage{scalefnt}
\usepackage[titles]{tocloft}
\usepackage{sectsty}
\allsectionsfont{\bf \scalefont{.7} \selectfont}
\subsectionfont{\bf \scalefont{.85} \it \selectfont}
\subsubsectionfont{\bf \scalefont{1} \it \selectfont}
\usepackage[T1]{fontenc}
\usepackage{lmodern}
\usepackage{bibspacing}

\def\blfootnote{\xdef\@thefnmark{}\@footnotetext}

\long\def\symbolfootnote[#1]#2{\begingroup%
\def\thefootnote{\fnsymbol{footnote}}\footnote[#1]{#2}\endgroup}

\makeatletter
\renewcommand{\@dotsep}{4.5}
\makeatother

\def\be{\begin{equation}}
\def\ee{\end{equation}}

\makeatletter
\def\@seccntformat#1{\csname the#1\endcsname.\quad}
\makeatother

\def\clock{{\count0=\time
           \divide\count0 60
           \ifnum\count0<10 0\fi\the\count0
           \multiply\count0 -60 \advance\count0 \time
           :\ifnum\count0<10 0\fi \the\count0
         }}
\newcommand{\timestamp}{{\small\vbox{\hbox{\tt\jobname.tex}
\hbox{\the\day/\the\month/\the\year, \clock}}}}

\def\FF{{\cal F}}

\def\II{{\cal I}}

\def\NN{{\cal N}}
\def\OO{{\cal O}}

\def\im{{\tt i}}

\newcommand{\beq}{\begin{equation}}
\newcommand{\eeq}{\end{equation}}
\newcommand{\ba}{\begin{array}}
\newcommand{\ea}{\end{array}}
\newcommand{\bea}{\begin{eqnarray}}
\newcommand{\eea}{\end{eqnarray}}

\newcommand{\Z}{\mathbb{Z}}
\newcommand{\C}{\mathbb{C}}

\newcommand{\N}{\mathbb{N}}

\newcommand{\tr}{\mathop{{\rm Tr}}}

\numberwithin{equation}{section}

\begin{document}

\begin{titlepage}
\begin{flushright}
CCTP-2011-10
\end{flushright}
\vskip 3.3cm
\begin{center}
\font\titlerm=cmr10 scaled\magstep4
    \font\titlei=cmmi10 scaled\magstep4
    \font\titleis=cmmi7 scaled\magstep4
    \centerline{\titlerm
    Comments on {\LARGE $F$}-maximization}
      \vspace{0.4cm}
     \centerline{\titlerm 
     and R-symmetry in 3D SCFTs }
\vskip 1.5cm
{\it Vasilis Niarchos}\\
\vskip 0.7cm
\medskip
{Crete Centre for Theoretical Physics,}\\ 
{Department of Physics, University of Crete, 71003, Greece}\\
\medskip
{niarchos@physics.uoc.gr}

\end{center}
\vskip .4in
\centerline{\bf Abstract}

\baselineskip 20pt
%

\vskip .5cm \noindent
We report preliminary results on the recently proposed $F$-maximization principle
in 3D SCFTs. We compute numerically in the large-$N$ limit the free energy on the 
three-sphere of an $\NN=2$ Chern-Simons-Matter theory with a single adjoint chiral 
superfield which is known to exhibit a pattern of accidental symmetries associated to
chiral superfields that hit the unitarity bound and become free. We observe that the
$F$-maximization principle produces a $U(1)$ R-symmetry consistent with previously
obtained bounds but inconsistent with a postulated Seiberg-like duality. Potential 
modifications of the principle associated to the decoupling fields do not appear to be 
sufficient to account for the observed violations.

\vfill
\noindent
March, 2011
\end{titlepage}\vfill\eject

\setcounter{equation}{0}

\pagestyle{empty}
\small
\tableofcontents
\normalsize
\pagestyle{plain}
\setcounter{page}{1}

\section{R-symmetry in three-dimensional $\NN=2$ SCFTs}
\label{problem}

Supersymmetric conformal field theories (SCFTs) with four real supersymmetries,
namely $\NN=2$ supersymmetry in three dimensions, have a conserved $U(1)$ R-symmetry
that sits in the same supermultiplet as the stress-energy tensor. In a general interacting 
SCFT this symmetry receives quantum corrections and the quantum numbers associated
with it become non-trivial functions of the parameters that define the theory. The computation
of the exact non-perturbative form of this symmetry is an important task with immediate 
implications, $e.g.$ one can deduce from it the scaling dimensions of the chiral ring operators.

In recent years considerable progress has been achieved in this problem in the context of 
four-dimensional $\NN=1$ SCFTs. Ref.\ \cite{Intriligator:2003jj} showed that the exact 
superconformal R-symmetry is the one that maximizes $a$ ---the coefficient of the Euler 
density in the conformal anomaly. 

More recently, evidence was presented \cite{Jafferis:2010un} 
that the free energy of the Euclidean CFT on a three-sphere
\beq
\label{probaa}
F=-\log|Z_{S^3}|
\eeq
plays a similar role in three-dimensional $\NN=2$ SCFTs. Putting the theory on a three-sphere
in a manner that preserves supersymmetry requires the introduction of extra couplings between
the matter fields and the curvature of $S^3$.  These couplings are determined by the choice of
a trial R-symmetry. In this way, $F$ is a function of the trial R-charges. It has been conjectured
\cite{Jafferis:2010un} that the exact $U(1)$ R-symmetry locally maximizes $F$.

So far, this proposal has passed a number of impressive tests 
\cite{Jafferis:2010un,Herzog:2010hf,Martelli:2011qj,Cheon:2011vi,Jafferis:2011zi,Amariti:2011hw}. 
It reproduces known perturbative results and agrees with expectations from the AdS/CFT 
correspondence. Currently, most of the analyzed examples refer to Chern-Simons-Matter quivers
with $\NN\geq 2$ supersymmetry and matter in the adjoint and bi-fundamental representation whose
free energy scales in the large 't Hooft limit as $N^{3/2}$. A potential discrepancy with expectations 
from the AdS/CFT correspondence was briefly reported for theories with chiral bifundamental fields 
in \cite{Jafferis:2011zi}.

More generally, it is natural to expect that the application of $F$-maximization 
will be subtle in cases with accidental global symmetries that are not visible in a weak coupling
formulation of the theory. Accidental symmetries can arise at strong coupling to modify the result.
Such situations are well known in four-dimensional $\NN=1$ SCFTs. For instance, in the case of 
accidental symmetries associated with fields hitting the unitarity bound it is known 
that the $a$-maximization principle should be modified appropriately to account for the decoupling
fields \cite{Kutasov:2003iy}. The validity (and possible modifications) of the proposed 
$F$-maximization principle in three dimensions has not been considered in such situations so far.

Our goal in this paper is to test the $F$-maximization principle in an $\NN=2$ Chern-Simons-Matter
(CSM) theory that is known to exhibit such strong coupling phenomena. The theory of interest is $U(N)$
$\NN=2$ Chern-Simons theory at level $k$ coupled to a single chiral superfield $X$ in the adjoint
representation. It is believed that this theory is superconformal for all values of $N,k$ 
\cite{Gaiotto:2007qi}. Moreover, one can argue that the exact R-charge of the superfield $X$ 
decreases towards zero as we make the theory more 
and more strongly coupled and that an increasing number of operators become free and decouple 
in this process. In fact, one can place specific non-perturbative constraints on how the exact R-charge 
decreases \cite{Niarchos:2008jb,Niarchos:2009aa}. The currently available information will be 
briefly reviewed in section \ref{hatA}. These constraints, which must be obeyed by the solution of any 
exact principle that determines the superconformal R-symmetry, like the proposed $F$-maximization, 
provide a novel way to check if the current formulation of $F$-maximization is valid, or in case it fails to 
detect how it fails and how it should be modified.

The partition function of $\NN=2$ SCFTs on $S^3$ reduces to a matrix integral
\cite{Kapustin:2009kz,Jafferis:2010un,Hama:2010av} via localization \cite{Pestun:2007rz}. 
In the case of the above single-adjoint CSM theory with gauge group $U(N)$ 
the matrix integral takes the form (up to an inconsequential overall factor)
\beq
\label{probab}
Z_{S^3}\sim  \int \left( \prod_{j=1}^N e^{\im \pi t_j^2 k} d t_j\right) \prod_{i<j}^N \sinh^2 (\pi (t_i-t_j))
\prod_{i,j=1}^N e^{\ell(1-R+\im (t_i-t_j))}
\eeq
where $\ell(z)$ is the function
\beq
\label{probac}
\ell(z)=-z \log \left(1-e^{2\pi \im z} \right)+\frac{\im}{2} \left[ \pi z^2+\frac{1}{\pi} {\rm Li}_2\left(
e^{2\pi \im z} \right) \right]-\frac{\im \pi}{12}\
~.
\eeq
$R$ is the trial R-charge over which we are instructed to maximize the free energy \eqref{probaa}.
We integrate over the $N\times N$ matrix eigenvalues $t_i$.

We will focus on the following 't Hooft limit of the theory
\beq
\label{probad}
N,k \to \infty~, ~~ \lambda=\frac{N}{k} = {\rm fixed}
~.
\eeq
In this limit the theory is parameterized by a single continuous parameter 
$\lambda \in [0,\infty)$.\footnote{We assume $k>0$. The case of negative $k$ can be obtained by
a simple parity transformation.} Accordingly, the exact R-charge is a function of 
$\lambda$.

In lack of a tractable analytic method, we will compute the matrix integral \eqref{probab} 
numerically in section \ref{Fmax} in the large-$N$ limit \eqref{probad} by solving a system 
of saddle point equations. Then we maximize the free energy to determine the exact 
R-symmetry. The result exhibits a function $R(\lambda)$ that decreases monotonically
towards zero as we increase the coupling $\lambda$ and remains in the vicinity of the lower bound
derived in \cite{Niarchos:2009aa} (see \eqref{hatAan}) without exhibiting any obvious
signs of violation. A different type of potential violation is noted, where the numerically
obtained behavior of the free energy appears to be inconsistent with the Seiberg-like duality
postulated in \cite{Niarchos:2008jb}. A more detailed discussion of this aspect will appear
elsewhere \cite{TN}.

The effects of decoupling operators modify the $F$-maximization principle. However, in this 
particular case such effects are subleading in $1/N$ and do not appear to be capable of 
producing numerically significant corrections at finite 't Hooft coupling $\lambda$. It is currently 
unclear whether it is possible to find a modification of the $F$-maximization principle that resolves 
the tension with the Seiberg-like duality.

The $F$-maximization matrix integral solution indicates the possibility
of a particular pattern of spontaneous supersymmetry breaking in the theories deformed by the 
superpotential interaction $W_{n+1}=\tr X^{n+1}$, $n=1,2,\ldots$; a pattern where the point 
of supersymmetry breaking is the same point where the operator $\tr X^{n+1}$ becomes a free 
operator in the undeformed theory. If correct, this pattern would imply that the exact superconformal 
R-charge is an oscillating (presumably monotonic) function that trails closely the curve 
$\frac{1}{2(1+\lambda)}$. The current numerical results partially verify this intuition. Relevant comments 
appear in sec.\ \ref{proposal}.

In the final section \ref{open} we conclude with a brief summary of the lessons of this work and
a list of interesting open problems.

\vspace{0.3cm}
\noindent
{\bf Note added.} In the first version of this paper an erroneous violation of the bounds
of Ref.\ \cite{Niarchos:2009aa} was reported due to a missing factor of 2 in the denominator
of the second term in eq.\ \eqref{Fmaxac}. In the current version this error has been corrected and
the numerical results updated. New comments related to a potential disagreement with a postulated
Seiberg-like duality have been added.

\section{${\bf \hat A}$ theory}
\label{hatA}

\vspace{-0.4cm}
\subsection{Definition and known facts}

We will focus on a particular class of three-dimensional $\NN=2$ SCFTs defined as 
$\NN=2$ Chern-Simons theory coupled to a single adjoint $\NN=2$ chiral superfield
$X$. Following \cite{Niarchos:2009aa} we will refer to this theory (in the
absence of superpotential deformations) as the ${\bf \hat A}$ theory.\footnote{The more 
general class of ${\bf \hat A}$ theories defined in \cite{Niarchos:2009aa} includes also $N_f$ 
pairs of (anti)fundamental chiral superfields $Q^i$, $\widetilde Q_i$. These theories are 
three-dimensional cousins of adjoint-SQCD in four dimensions.} The ${\bf \hat A}$ 
theory is characterized by two integers: the rank of the gauge group $G$ (we will take
$G=U(N)$ in this work), and the level of the Chern-Simons interaction $k$, which is also an
integer. It is believed that this theory is exactly superconformal at the quantum level 
for any values of $N$, $k$ \cite{Gaiotto:2007qi}.

In the large-$N$ limit \eqref{probad} there is a single continuous parameter, the 't Hooft 
coupling $\lambda=N/k$. The theory is weakly coupled when $\lambda \ll 1$, in which case
we can treat it with standard perturbative techniques. In this regime the superconformal
R-charge behaves as \cite{Gaiotto:2007qi}
\beq
\label{hatAaa}
R(\lambda)=\frac{1}{2}-2\lambda^2+\OO(\lambda^4)
~.
\eeq
In this note we are interested in the exact non-perturbative version of \eqref{hatAaa}.

There is no known holographic description of the ${\bf \hat A}$ theory. Ref.\ \cite{Gaiotto:2007qi}
explored the possibility of a holographic description in M-theory based on $N$ M5-branes 
wrapping a special Lagrangian Lens space $S^3/\Z_k$. No AdS$_4$ solution was found for 
this system in supergravity which agrees
with the expectation that $\alpha'$ corrections will be important in a type IIA dual string theory
formulation of this theory. Hence, in this case we cannot invoke the AdS/CFT correspondence 
to gain information about the strong coupling behavior of the R-symmetry.

Instead, it is possible to obtain useful information about the exact R-symmetry by analyzing the
properties of the theory under the superpotential deformations 
\beq
\label{hatAab}
W_{n+1}=\frac{g_{n+1}}{n+1} \tr X^{n+1}~,~~ n=1,2,\ldots
~.
\eeq
As we review in a moment, there are regimes along the $\lambda$-line where these interactions
are relevant and drive the theory to a new interacting IR fixed point. We will refer to the theory 
deformed by the chiral operator $\tr X^{n+1}$ as the ${\bf A}_{n+1}$ theory. An argument based
on a D-brane realization of this theory in string theory \cite{Niarchos:2008jb} shows that 
\begin{itemize}
\item[$(i)$] The superpotential deformation $W_{n+1}$ lifts the supersymmetric vacuum for
\beq
\label{hatAac}
N > nk~~~~({\rm equivalently ~in~the~large-}N{\rm~limit:}~~ \lambda> n)
~.
\eeq  
\item[$(ii)$] The theory exhibits a Seiberg-like duality. 
The $U(N)$ theory at level $k$ and superpotential
deformation $W_{n+1}$ is dual to the $U(nk-N)$ theory at the same level $k$ and superpotential
$W_{n+1}$. In the large-$N$ 't Hooft limit the duality acts by taking 
\beq
\label{hatAad}
\lambda \to n-\lambda
~.
\eeq 
\end{itemize}

This picture has important implications for the R-symmetry in the undeformed theory ${\bf \hat A}$
\cite{Niarchos:2009aa}.
Classically, $i.e.$ at $\lambda \ll 1$, the chiral operators $\tr X^{n+1}$ ($n=4,5,\ldots$) are all
irrelevant and become more irrelevant the larger $n$ is. The fact that there
are values of $\lambda \in \N$ for which any operator $\tr X^{n+1}$ can lift the supersymmetric 
vacuum (no matter how large $n$ is) implies that the exact R-charge decreases as we increase
$\lambda$ and eventually tends to zero at infinite $\lambda$. 

More specifically, it implies that there
is a sequence of critical couplings $\lambda^*_{n+1}$ such that
\beq
\label{hatAae}
0=\lambda^*_2=\lambda^*_3=\lambda^*_4 < \lambda^*_5<\cdots< \lambda^*_n<\lambda^*_{n+1}
<\cdots
\eeq
where each time one of the chiral operators $\tr X^{n+1}$ becomes marginal. By definition, 
$\lambda^*_{n+1}$ is the value of the 't Hooft coupling where the operator $\tr X^{n+1}$ has
scaling dimension
\beq
\label{hatAaf}
\Delta(\tr X^{n+1})=(n+1) R(\lambda^*_{n+1})=2 ~~ \Leftrightarrow ~~
R(\lambda^*_{n+1})=\frac{2}{n+1}
~.
\eeq
Clearly, the generic operator $\tr X^{n+1}$ must become marginal before it becomes capable to 
lift the supersymmetric vacuum at $\lambda_{n+1}^{\rm SUSY}=n$. This implies 
\beq
\label{hatAag}
\lambda_{n+1}^*<n
\eeq
and 
\beq
\label{hatAai}
\Delta(\tr X^{n+1})|_{\lambda=n}=(n+1)R(n)<2 ~~\Leftrightarrow ~~
R(n)<\frac{2}{n+1}
~.
\eeq

A more strict upper bound on $\lambda^*_{n+1}$ can be deduced from the Seiberg-like duality
\eqref{hatAad}. Requiring the existence of a finite range of $\lambda$-values within which 
the deforming operator $\tr X^{n+1}$ is relevant both in the $U(N)$ theory and its $U(nk-N)$ dual
implies
\beq
\label{hatAana}
\lambda_{n+1}^* < n-\lambda^*_{n+1} ~~\Leftrightarrow ~~ \lambda^*_{n+1}<\frac{n}{2}
~.
\eeq
$n-\lambda^*_{n+1}$ is the point where $\tr X^{n+1}$ becomes marginal in the dual theory.
The interval $[\lambda^*_{n+1}, n-\lambda^*_{n+1}]$ plays in the ${\bf A}_{n+1}$ 
theory the analog of the standard conformal window in 4d SQCD. The point $\lambda=\frac{n}{2}$
is a self-dual point for Seiberg duality in the ${\bf A}_{n+1}$ theory.

As we increase $\lambda$ beyond some $\lambda^*_{n+1}$ we reach the critical coupling 
$\lambda^*_{n'+1}$ $(n'>n)$ of another operator $\tr X^{n'+1}$. It so happens that there is an
integer $n'$ for which $\tr X^{n'+1}$ is marginal and simultaneously $\tr X^{n+1}$ hits the 
unitarity bound and becomes free. This occurs precisely when
\beq
\label{hatAaj}
\Delta(\tr X^{n+1})=(n+1)R(\lambda^*_{n'+1})=\frac{2(n+1)}{n'+1}=2 ~~\Leftrightarrow~~
n'=4n+3
~.
\eeq
Once we reach $\lambda^*_{4(n+1)}$, where the operator $\tr X^{n+1}$ becomes free, we 
cannot use it any longer to deform the ${\bf \hat A}$ theory without destabilizing the supersymmetric
vacuum. Hence, the spontaneous supersymmetry breaking point $\lambda^{\rm SUSY}_{n+1}=n$
of the ${\bf A}_{n+1}$ theory cannot be greater than $\lambda^*_{4(n+1)}$. That implies a further pair
of inequalities \cite{Niarchos:2009aa}
\beq
\label{hatAak}
n\leq \lambda^*_{4(n+1)}
~,
\eeq
and
\beq
\label{hatAal}
\Delta(\tr X^{n+1})|_{\lambda=n}=(n+1) R(n)\geq \frac{1}{2} ~~\Leftrightarrow ~~
R(n) \geq \frac{1}{2(n+1)}
~.
\eeq

Combining the inequalities \eqref{hatAana}, \eqref{hatAai}, \eqref{hatAak}, \eqref{hatAal} we find
\beq
\label{hatAam}
\left[ \frac{n-3}{4} \right]\leq \lambda^*_{n+1}<\frac{n}{2}
~,
\eeq
and
\beq
\label{hatAan}
\frac{1}{2(n+1)}\leq R(n) < \frac{2}{n+1}~, ~~ n=1,2,\ldots
~.
\eeq
Assuming $R(\lambda)$ is a monotonically decreasing function,\footnote{The intuition that
gauge interactions work to decrease the R-charge with increasing $\lambda$ makes this
assumption plausible. However, it is far from obvious that this is a correct statement in the 
exact theory. We will see that the R-charge derived from $F$-maximization satisfies this property
of monotonicity. We stress that the validity of the inequalities \eqref{hatAam}, \eqref{hatAan}
does not rely on this assumption.} the improved upper bound on 
$\lambda^*_{n+1}$ in \eqref{hatAam} further implies
\beq
\label{hatAao}
R(\lambda)<\frac{2}{2\lambda+1}~, ~~{\rm for}~~ \lambda=\frac{n}{2} ~,~~ n=1,2,\ldots
~.
\eeq

In the next section we will test whether $F$-maximization in its current form obeys these inequalities.

To summarize, in the ${\bf \hat A}$ theory we encounter the following situation. At weak coupling
the operator $\tr X$ is free and decoupled but all the other chiral ring operators $\tr X^{n+1}$ ($n>0$)
are interacting. As we increase $\lambda$ more and more of the operators from the chiral ring are
decommissioned. For any $n$ there is always a value of $\lambda$ above which the scaling dimension
of the operator $\tr X^{n+1}$ takes the free field value $\frac{1}{2}$. According to standard lore, the
operator becomes a free field at that point and decouples from the rest of the theory. Hence, with
increasing $\lambda$ more and more of the bottom part of the chiral ring decouples.

\subsection{A brief note on moduli spaces}

Before moving to the computation of the free energy on the three-sphere let us interject a 
comment on the deformation $W_{n+1}$ that gives rise to the theories ${\bf A}_{n+1}$.

By definition, the operator $\tr X^{n+1}$ is marginal at $\lambda=\lambda^*_{n+1}$.
We would like to ask: is the superpotential deformation $W_{n+1}$ an exactly marginal 
deformation at  $\lambda^*_{n+1}$?

The technology of Ref.\ \cite{Green:2010da} allows us to give a definite answer to this question.
The superpotential deformation \eqref{hatAab} breaks the global $U(1)_X$ symmetry that rotates
the superfield $X$ and gives a non-vanishing $D^a$ (in the language of \cite{Green:2010da}).
Hence, $g_{n+1}$ at $\lambda=\lambda^*_{n+1}$ is a marginally-irrelevant coupling. This is
similar to what happens with classically marginal superpotential deformations in Wess-Zumino 
models. 

As we increase $\lambda$ above $\lambda^*_{n+1}$, the superpotential deformation 
\eqref{hatAab} becomes relevant and there is a flow towards a fixed point (the superconformal
field theory we denote by ${\bf A}_{n+1}$). At this point the superpotential coupling $g_{n+1}$  
becomes a function of $\lambda$
\beq
\label{moduliab}
g=g_{n+1}(\lambda)
~.
\eeq
This can be shown explicitly in conformal perturbation theory when 
$\lambda-\lambda^*_{n+1}\ll 1$(see $e.g.$ \cite{Green:2010da}).
Intuitively, an IR fixed point arises from a balancing of two counteracting sources: the gauge
interactions that work to decrease the R-charges and the superpotential interactions that
work to increase them.

\section{The partition function on $S^3$ and $F$-maximization in the large-$N$ limit}
\label{Fmax}

We proceed to compute the matrix integral \eqref{probab} in the large-$N$ limit \eqref{probad}, 
implement the $F$-maximization principle and determine $R$ as a function of $\lambda$. 
Expressed as a function of $N, \lambda, R$ the partition function $Z_{S^3}$ reads
\beq
\label{Fmaxaa}
Z_{S^3}=\int  \left( \prod_{j=1}^N e^{\frac{\im \pi N}{\lambda} t_j^2} 
d t_j\right) \prod_{i<j}^N \sinh^2 (\pi (t_i-t_j))
\prod_{i,j=1}^N e^{\ell(1-R+\im (t_i-t_j))}
=e^{-\FF(N,\lambda, R)}
~.
\eeq
The free energy that we want to maximize with respect to $R$ is
\beq
\label{Fmaxab}
F=\frac{1}{2}(\FF+\bar \FF)
~.
\eeq

In the large-$N$ limit the main contribution to $Z_{S^3}$ comes from saddle point configurations
that obey the system of algebraic equations
\beq
\label{Fmaxac}
\II_i \equiv \frac{\im}{\lambda}t_i+\frac{1}{N} \sum_{j\neq i}^N \left[ \coth(\pi t_{ij})
-\frac{(1-R)\sinh(2\pi t_{ij})+t_{ij}\sin(2\pi R)}{\cosh(2\pi t_{ij})-\cos(2\pi R)}\right]=0
~, ~~ i=1,2,\ldots,N
\eeq
where we have defined
\beq
\label{Fmaxad}
t_{ij}=t_i-t_j
~.
\eeq
In general, the $N$ $t_i$'s that solve these equations are complex numbers. In this case,
they are $\C$-valued functions of the parameters $R,\lambda$.

At a saddle point configuration
\beq
\label{Fmaxada}
-\FF(\lambda, N)=\sum_{i=1}^N \frac{\im \pi N}{\lambda} t_i^2
+\sum_{i<j}^N \log \sinh^2\left(\pi t_{ij}\right)
+\sum_{i,j=1}^N \ell \left(1-R+\im t_{ij} \right)
~.
\eeq

We are not aware of an efficient analytic method of solution of these equations in this particular
case, so we will proceed with a more elementary numerical approach. As pointed out in 
\cite{Herzog:2010hf} it is convenient to view such equations as equations describing 
the equilibrium configuration of $N$ point particles whose coordinates are given by the 
complex numbers $t_i$. The equilibrium configuration can be found by introducing a fictitious
time coordinate $\tau$ and considering the dynamical evolution described by the differential
equation
\beq
\label{Fmaxae}
a \frac{d t_i}{d\tau}=\II_i
~.
\eeq
By suitably choosing the constant $a$ we can arrange for solutions that converge very quickly
to an equilibrium configuration described by the system \eqref{Fmaxac}.

\begin{figure}[t!]
\centering
\includegraphics[height=5.3cm]{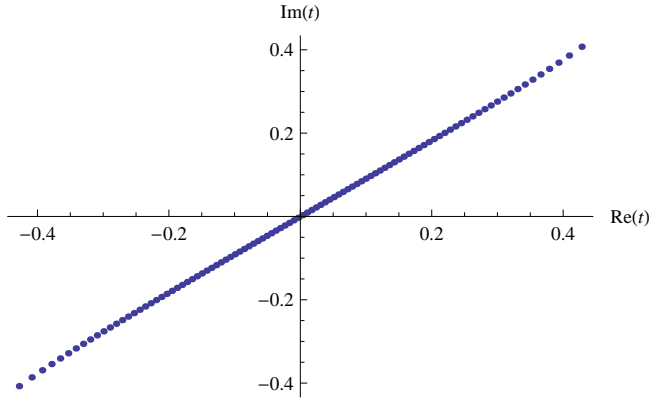}
\bf\caption{\it \small A typical distribution of the eigenvalues $t_i$ $(i=1,\ldots, N)$ in the
complex plane. This particular plot has been obtained for $N=100$, $\lambda=1$ and 
$R=0.28$.}
\label{eigen}
\end{figure}

We implemented this approach on the computer (with the use of {\sc mathematica}) for a 
wide range of $\lambda$ and $R$ values. As a typical value of large $N$ we used $N=100$.
An appropriate choice of $a$ is $a=-\im$.

A typical saddle point configuration appears in Fig.\ \ref{eigen}. The configuration is symmetric under
the transformation $t_i \to -t_i$ as is evident from the equations \eqref{Fmaxac}.
The specific arrangement of the eigenvalues in the complex plane depends on the values of 
$\lambda$ and $R$. However, in general, we observe that the eigenvalues are oriented along 
a line of approximately 45 degrees and that the size of their domain increases with increasing 
$\lambda$ (at fixed $N$ and $R$).  

Maximizing the free energy \eqref{Fmaxab} with respect to $R$ we determine the R-charge as
a function of $\lambda$. The result of this calculation is plotted in Figs.\ \ref{Rweak}, \ref{Rstrong}.
Let us discuss separately the regimes with $\lambda$ of order one and $\lambda\gg 1$.

\subsection{The R-charge at $\lambda\sim \OO(1)$}
\label{Ro1}

\begin{figure}[t!]
\centering
\includegraphics[height=5.3cm]{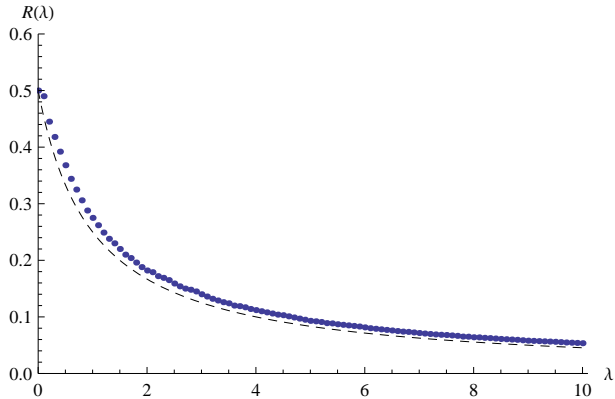}
\bf\caption{\it \small The numerically computed R-charge curve in the regime of 
$\lambda\sim \OO(1)$.
The dashed curve represents the lower bounding function $\frac{1}{2(1+\lambda)}$.}
\label{Rweak}
\end{figure}

Fig.\ \ref{Rweak} zooms into the region of interest.
This region includes the perturbative regime of $\lambda\ll 1$, but also a regime of 
strong coupling at $\lambda$ of order 1 (we present data up to $\lambda=10$). 
In this regime we find that the free energy $F$ scales with $N$ as $\OO(N^2)$. 

At very weak coupling the curve follows very closely the result of the perturbative calculation 
\eqref{hatAaa}. The successful matching of the perturbative result with the result obtained from 
$F$-maximization was noticed already in \cite{Jafferis:2010un} for gauge group $SU(2)$ and 
more recently in \cite{Amariti:2011hw} for the general $SU(N)$ case. In this paper we are 
considering the case of $U(N)$ gauge group. It is not difficult to show, using a trick in 
\cite{Amariti:2011hw}, that the $SU(N)$ and $U(N)$ versions of the matrix integral \eqref{Fmaxaa} 
are simply related by the equation
\beq
\label{SUvsU}
Z_{S^3}[SU(N)]=\frac{1}{\sqrt{i\lambda}}e^{-\ell(1-R)} Z_{S^3}[U(N)]
~.
\eeq
Hence, to leading order in the planar limit, $F$-maximization leads to the perturbative
result \eqref{hatAaa} in the $U(N)$ case as well in agreement with expectations.

Away from the perturbative regime we observe the R-charge decreasing 
monotonically.\footnote{We have checked (for the saddle point solutions reported in this paper) 
that $F$ has a single extremum in the regime of this subsection.}
It remains well below the two upper bounds and close, but above, the lower bound set by
the curve $\frac{1}{2(1+\lambda)}$. Recall that this curve places a lower bound on $R$ only 
when $\lambda$ is an integer. A list of the numerically determined values of $R$ at 
$\lambda=1,2,\ldots,10$ and the corresponding lower bounding values appears in Table \ref{list}.
As we increase $\lambda=n\in \N$ the difference $R_{num}(n)-R_{bound}(n)$ decreases but 
remains positive respecting the bound \eqref{hatAan}. By analyzing the $N$-dependence of 
the numerical results we find that the typical error is of the order of a few percent. The difference 
$R_{num}-R_{bound}$ in Table \ref{list} is of the order of 10\%.

\begin{table}[t!]
\centering
\begin{tabular}{|c||c|c|c|c|c|c|c|c|c|c|} \hline
 $\lambda$ & 1 & 2 & 3 & 4 & 5 & 6 & 7 & 8 & 9 & 10 \\ 
 \hline
 $R_{num}$ & 0.275 & 0.182 & 0.14 & 0.112 & 0.093 & 0.0815 & 0.071 & 0.064 & 0.058 & 0.0535 \\
 \hline
 $R_{bound}$ & 0.25 & 0.1667 & 0.125 & 0.1 & 0.083 & 0.0714 & 0.0625 & 0.0556 & 0.05 & 0.0455 \\
 \hline 
\end{tabular}
\bf\caption{\it \small Numerically determined values of $R$ at $\lambda=1,2,\ldots,10$ and the 
corresponding lower bounding values.}
\label{list}
\end{table}

The first chiral operator that saturates the unitarity bound is $\tr X$. As is evident already from the 
perturbative result \eqref{hatAaa} $\tr X$ is a free operator at any value of $\lambda$.

Non-perturbatively we observe that the second operator that hits the unitarity bound is 
$\tr X^2$. That occurs very close to $\lambda=1$, equivalently 
\beq
\label{Fmaxba}
\Delta\left(\tr X^2 \right)\Big |_{\lambda=1}-\frac{1}{2}=2R(1)-\frac{1}{2}\sim 0.025
~.
\eeq
As we increase $\lambda$ more and more single-trace operators decouple sequentially.
As we reviewed in section \ref{hatA}, $\lambda=n\in \Z$ is a special coupling in the 
${\bf A}_{n+1}$ theory. For example, the ${\bf A}_2$ theory is the mass-deformed 
${\bf \hat A}$ theory. In that case, in the far infrared the deformed theory flows to 
the $\NN=2$ Chern-Simons theory, a topological theory known to exhibit spontaneous 
supersymmetry breaking at $\lambda > 1$ \cite{Witten:1999ds}. Here we observe 
numerically that the deforming operators $\tr X^{n+1}$ come very close to becoming free as we 
approach the supersymmetry breaking point of the ${\bf A}_{n+1}$ theory. We will return to this 
point in section \ref{proposal}.

\subsection{The R-charge at $\lambda\gg 1$}

\begin{figure}[t!]
\centering
\includegraphics[height=5.3cm]{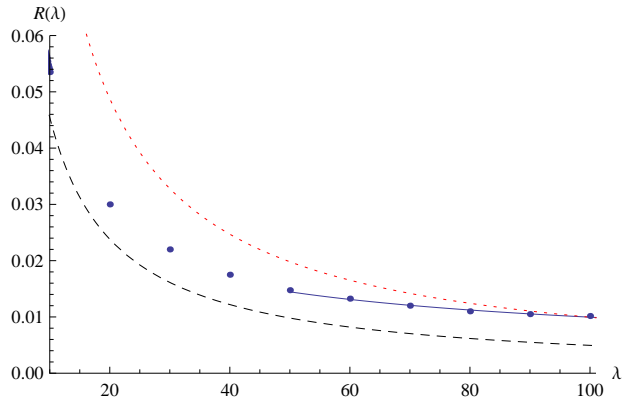}
\bf\caption{\it \small Numerical results in the regime of very strong 't Hooft coupling.
The black and red curves represent the lower bounding function $\frac{1}{2(1+\lambda)}$
and the upper bounding function $\frac{2}{2\lambda+1}$ respectively.}
\label{Rstrong}
\end{figure}

As we move into the regime of stronger and stronger 't Hooft coupling the numerically
determined R-charge curve continues to behave in a monotonically decreasing fashion
approaching zero. The numerical computation becomes slower and the $N$-dependence
increases. In addition, order-$[\lambda]$ operators are decoupled. Accordingly, the effect of this 
decoupling is expected to increase as we increase $\lambda$.
 
As an indicative illustration of the solutions of $F$-maximization in this regime
we exhibit in Fig.\ \ref{Rstrong} numerical results for $N=100$ and $\lambda$ up to $N$.
A fit of the data for $\lambda$ above 50 provides the following estimate for the asymptotic behavior
\beq 
\label{estim}
R(\lambda)\sim 0.119 ~ \lambda^{-0.538}
~.
\eeq
The numerical curve appears to cross the upper bounding curve $\frac{2}{2\lambda+1}$
in the vicinity of $\lambda\sim 100$. In this paper we will refrain from drawing specific conclusions
from this result and postpone a more detailed exploration of this regime to a future publication. 

At the same time, we observe that the free energy continues to scale like $N^2$. In the language of 
\cite{Jafferis:2011zi} this scaling is due to the non-cancelation of long-range forces on the 
eigenvalues in the saddle point approximation.

\subsection{Comments on a potential discrepancy}

The above application of the $F$-maximization recipe does not produce any clear violations
of the bounds of section \ref{hatA} in the regime of subsection \ref{Ro1}. In principle, when
some operators become free and decouple from the rest of theory new accidental symmetries
occur that can mix with the R-symmetry. In such cases, the exact R-symmetry that we want to 
determine does not refer to the decoupled operators anymore and any extremization principle 
should properly take this fact into account by `subtracting' the free operators. Similar modifications 
of the $a$-maximization principle in four-dimensional SCFTs have been discussed in 
\cite{Kutasov:2003iy}. In that case, one is instructed to maximize a modified $a$-function where 
the 't Hooft anomalies associated to the decoupled fields have been subtracted. Clearly, a 
similar `subtraction' recipe must be implemented in the case of $F$-maximization. 

In the regime of subsection \ref{Ro1} only a finite (order-one) number of single-trace operators
decouples. Hence, in the absence of other effects, such `subtraction' modifications are expected
to have negligible effects to the free energy (an order-$N^2$ quantity) and $F$-maximization 
should go through unobstructed. From this point of view, it is good news that no violations
of the bounds of section \ref{hatA} are observed in this regime. On a related note, it is interesting
to compare the current situation with the corresponding situation in adjoint-SQCD in four 
dimensions \cite{Kutasov:2003iy}. In that case, the fields whose decoupling produces a sizable 
effect in the $a$-maximization procedure in the large-$N$ limit are mesonic fields whose number 
is order-$N^2$, a number scaling similarly to the free energy. 

The non-perturbative bounds reviewed in section \ref{hatA} is only part of the picture expected 
to describe the ${\bf \hat A}$ and ${\bf A}_{n+1}$ theories. Another important part of the story is
the Seiberg-like duality in the ${\bf A}_{n+1}$ theories proposed in \cite{Niarchos:2008jb} (where
also several checks of the duality were performed). In the large-$N$ limit this duality maps the 
${\bf A}_{n+1}$ theory 
at coupling $\lambda$ to the ${\bf A}_{n+1}$ theory at coupling $n-\lambda$. This duality has 
several non-trivial consequences. One of them is the prediction that inside the `conformal window' 
$(\lambda^*_{n+1},n-\lambda^*_{n+1})$ the free energies of the dual theories at $\lambda$ and
$n-\lambda$ should match. Since localization is blind to the superpotential interactions, 
we can compute these free energies with no extra effort simply by setting $R$ in the previous
matrix integral computation to the value dictated by the requirement that the superpotential 
interaction is marginal (in the case of the ${\bf A}_{n+1}$ theory that is $R=\frac{2}{n+1}$).
Then, the Seiberg-like duality predicts the following set of equations
\beq
\label{dualityaa}
\FF\left(\frac{2}{n+1},\lambda\right)=\FF \left( \frac{2}{n+1},n-\lambda\right)~~, ~~ 
\lambda\in (\lambda^*_{n+1},n-\lambda^*_{n+1})~, ~~ n=3,4,\ldots
\eeq
which allow us to probe the validity of the matrix integral \eqref{probab} (and $F$-maximization)
over a wider range of $R$ and $\lambda$ values.

Preliminary results based on the solutions of the saddle point equations presented above 
show that the oscillatory behavior implied by the equations \eqref{dualityaa} is not observed.
Assuming the validity of the duality, this implies the presence of additional effects above
$\lambda\sim 1$ which have to be taken properly into account in order to make sense of 
$F$-maximization in that regime. A detailed discussion of these issues will be presented in 
\cite{TN}.

\section{Insights into the exact R-symmetry of the ${\bf \hat A}$ theory}
\label{proposal}

In anticipation of a modified $F$-maximization principle that 
gives the exact R-symmetry at arbitrary 't Hooft coupling we would like to 
offer in this section a few tentative comments on a possible result.
We warn the reader that some of the statements that follow are speculative and
we currently have no independent conclusive means to check whether this scenario is 
actually realized by the exact R-symmetry.

In subsection \ref{Ro1} we observed that the R-charge obtained from $F$-maximization
remains in the vicinity of the curve $\frac{1}{2(1+\lambda)}$. The current formulation of 
$F$-maximization seems to be trustable at least within the regime $\lambda \in [0,1]$, 
where it reproduces the perturbative result correctly and there are no obvious discrepancies
with known or expected facts. We would like to suggest that the proximity of the R-charge 
curve to the bounding curve $\frac{1}{2(1+\lambda)}$ remains true at arbitrary values of 
$\lambda$. 

Part of our intuition about this property comes from a qualitatively similar theory in four
dimensions, $SU(N_c)$ SQCD with $N_f$ fundamental/antifundamental superfield pairs and 
a single adjoint superfield. This theory was analyzed with $a$-maximization techniques in 
\cite{Kutasov:2003iy}. In the large-$N$ limit, there is a single parameter that controls the 
dynamics, the ratio $x=\frac{N_c}{N_f}$. This theory, which is asymptotically free for 
$x>\frac{1}{2}$, is believed to flow in the IR to a non-trivial fixed point. When deformed by 
the superpotential $W_{n+1}\sim \tr X^{n+1}$ one finds that there is a supersymmetric vacuum
only when $x<n$. The exact R-symmetry can be determined with $a$-maximization, 
properly defined to account for decoupling fields. In particular, the R-charge of the adjoint 
superfield is found be a monotonically decreasing function of $x$ that approaches zero at 
large $x$ with the asymptotics
\beq
\label{proposalaa}
R(x)\sim \frac{4-\sqrt{3}}{3x}
~.
\eeq
Hence, at $x=n\in \N$ the corresponding asymptotics of the scaling dimension of the generic 
single trace operator $\tr X^{n+1}$ obeys the relation
\beq
\label{proposalab}
\Delta\left( \tr X^{n+1}\right)\Big |_{x=n}-1=\frac{3}{2}(n+1)R(n)-1\sim 1-\frac{\sqrt 3}{2}\sim 0.13
~.
\eeq
The operator $\tr X^{n+1}$ comes close to becoming a free field
at the point where the supersymmetric vacuum is lifted in the deformed theory and the R-charge
curve trails the lower bound curve $\frac{2}{3(1+\lambda)}$ that follows from the inequality
\beq
\label{proposalac}
\frac{2}{3(1+n)}<R(n)~, ~~ n\in \N
~.
\eeq

Of course, there are important differences between the  three-dimensional CSM theory that 
we are discussing here and the above four-dimensional adjoint-SQCD theory. For example, the 
latter has necessarily a non-vanishing number of fundamentals. Despite this fact, we propose 
that it is not unreasonable to anticipate some qualitative similarities (indeed, we already observe
several non-trivial similarities, $e.g.$ similarities in the supersymmetry breaking pattern).

Furthermore, given the small numerical value on the rhs of eq.\ \eqref{Fmaxba}, we would like 
to take the above picture one step further and suggest the possibility of the following property of 
the exact R-symmetry\footnote{The numerical results of the previous section suggest that these 
equalities are approximately but not exactly correct (at least for small enough $\lambda$). 
Since the deviation from \eqref{proposalad} is of a few percent, and thus comparable with the 
expected numerical error, we will not use the numerical results here to make a definite conclusive 
statement about the fate of the equalities \eqref{proposalad} and will instead proceed to explore 
their implications.} 
\beq
\label{proposalad}
\Delta \left(\tr X^{n+1} \right)\Big |_{\lambda=n}=\frac{1}{2} ~~
\Leftrightarrow ~~ R(n)=\frac{1}{2(1+n)}
~.
\eeq
If correct, this property implies that the deformation $W_{n+1}\sim \tr X^{n+1}$ yields an IR 
fixed point with a supersymmetric vacuum all the way up to the point where the deforming
operator becomes free. Since we have a theory of a single chiral superfield in this case,
this property may not be unnatural. It seems less likely to be exact in situations that involve 
the dynamics of extra fields, for example extra fields in the fundamental representation.

Assuming the validity of \eqref{proposalad} as a working hypothesis we arrive at the following
picture for the $\lambda$-dependence of the exact R-symmetry of the ${\bf \hat A}$ theory.

In principle, the exact R-charge curve can exhibit one of the following three different types
of dependence on $\lambda$:
\begin{itemize}
\item[$(i)$] $R(\lambda)$ oscillates indefinitely in the vicinity of the function
\beq 
\label{proposalae}
f(\lambda)=\frac{1}{2(1+\lambda)}
\eeq
passing through the points \eqref{proposalad} at $\lambda=n\in \N$.
\item[$(ii)$] $R(\lambda)$ coincides with the function $f(\lambda)$.
\item[$(iii)$] $R(\lambda)$ oscillates in a finite interval of $\lambda$ and coincides with the
function $f(\lambda)$ in its complement.
\end{itemize}

The only feasible possibility is possibility $(i)$. Possibility $(ii)$ is excluded immediately by the
perturbative result \eqref{hatAaa}. Possibility $(iii)$ is excluded by the invariance of $R$ 
under the transformation $k\to -k$, or equivalently $\lambda \to -\lambda$. This transformation
can also be seen as a parity transformation. Although this is not a symmetry of the theory, it is
a symmetry of the spectrum and hence a symmetry of the R-charge.

Hence, under the assumption \eqref{proposalad}, we conclude that the R-charge is an oscillating
(presumably monotonic) function of $\lambda$ with the following asymptotics at strong coupling
\beq
\label{proposalaf} 
R(\lambda)\sim \frac{1}{2\lambda}
~.
\eeq
It will be interesting to verify how close to reality the above picture is.

\section{Open problems}
\label{open}

We argued that the application of $F$-maximization is a subtle exercise  
in a three-dimensional SCFT with accidental symmetries associated 
with fields that reach the unitarity bound and decouple. In a regime where the 
effects of the decoupling fields appear to be negligible we found evidence that
$F$-maximization respects the non-perturbative bounds of \cite{Niarchos:2009aa}
but fails to reproduce results consistent with the Seiberg-like duality of \cite{Niarchos:2008jb}.
Since our arguments were based solely on a numerical computation it would be useful to 
substantiate them further with additional analytic evidence.

The main remaining tasks are:
$(i)$ to determine conclusively if and how the current application of $F$-maximization fails, 
and resolve the puzzles that have emerged, and $(ii)$ if modifications are needed to determine 
them and obtain results consistent with known facts. If $(ii)$ can be successfully 
implemented, it is interesting to verify or disprove whether the exact $R$-symmetry 
behaves in the manner anticipated in section \ref{proposal}. Our ultimate hope is to
obtain lessons that are applicable beyond the specific theory that was discussed in this
paper.

For example, it would be very interesting to examine the corresponding properties of the more general
class of ${\bf \hat A}$ theories in \cite{Niarchos:2009aa} that include $N_f$ additional pairs of
fundamental/antifundamental superfields $Q^i,\widetilde Q_i$. In these theories there are two
unknown R-charges, $R(Q)$ and $R(X)$, which are non-trivial functions of the parameters
$x=\frac{N_f}{N}$ and $\lambda=\frac{N}{k}$ (in the large-$N$, $N_f$ limits). It is known 
\cite{Niarchos:2009aa} that $R(X)$ decreases at strong coupling towards a non-zero value
$R(X)_{\rm lim}>\frac{1}{2([x]+2)}$. No particular information is currently available for $R(Q)$
beyond the perturbative regime. It is possible that as we increase the 't Hooft coupling some
meson-like operators hit the unitarity bound and decouple. In that sense, this particular example 
may prove more appropriate in studying modifications of the $F$-maximization principle associated
to decoupling fields.

Finally, a related issue has to do with the postulated $F$-theorem in \cite{Jafferis:2011zi}, which 
states that the free energy $F$ on the three-sphere \eqref{probaa} decreases along RG flows and
plays the role of a $c$-function in three dimensions. Potential modifications of the 
$F$-maximization principle may have direct implications to the formulation of 
an $F$-theorem as well.

\section*{Acknowledgements}

I would like to thank Elias Kiritsis and Dario Martelli for helpful comments and 
discussions. I am also grateful to David Kutasov for helpful discussions and correspondence, 
Daniel Jafferis for explaining a point in Ref.\ \cite{Jafferis:2011zi} and Takeshi Morita for the 
collaboration in a related project. This work was partially supported by the European Union grants 
FP7-REGPOT-2008-1-CreteHEPCosmo-228644 and PERG07-GA-2010-268246.





\end{document}